\documentclass[11pt]{article}

\usepackage{ebgaramond}
\usepackage[svgnames]{xcolor}
\usepackage[super]{nth}
\usepackage{fullpage}
\usepackage{parskip}
\usepackage{natbib}
\usepackage{hyperref}
 
\hypersetup{
    colorlinks = True,
    allcolors =  Crimson
}    

\title{Agentic research is oxymoronic}

\author{Natalie B. Hogg\thanks{Institute of Astronomy \& Kavli Institute for Cosmology, University of Cambridge. \href{mailto:nbh25@cam.ac.uk}{nbh25@cam.ac.uk}.
}}

\date{\nth{31} August 2026}

\begin{document}

\maketitle

\begin{abstract}
\noindent The use of agentic large language models
obviates human interpretation of scientific
results, and will lead to substantial distrust in
the literature.
\end{abstract}

Astronomy has recently found a new problem to chew over: the use of large language models (LLMs) in scientific practice. Not so long ago, I was a vocal proponent of the `ban-and-punish' stance when it came to LLMs. I believed that LLMs were a danger to the environment, to creativity and to our science. Then, eight months ago, and in spite of this position, I began to use LLMs on a daily basis in my work. The push factor for me was starting a new job, and joining the \href{https://handley-lab.co.uk/}{Handley Research Group} -- in which LLM use is well in advance of what the rest of astronomy is doing. In particular,
members of the Handley Lab use LLMs in agentic mode, where the model is given access to the user's computer and can read files, execute terminal commands and run code, in some cases creating and executing a plan of work completely autonomously, writing up its findings at the end. I found this impressive, and was curious to try it for myself.

I undoubtedly felt peer-pressure from my new environment, not to mention the feeling that `AI is the future' -- the narrative of inevitability about these tools which is unsubtly pushed by the multi-billion dollar technology firms that own the models with which most people currently interact. I felt that I ought to start using agentic AI so as not to fall behind my peers. Modern-day academia has delighted in breeding the miasmatic `publish or perish' environment -- could agentic AI be a way to work faster, and thus to stay ahead of the game?

Much public and private discourse has already occurred on this topic~\citep{Trotta2025, DHogg2026, Peiris2026, Bertone2026}, but it is mostly being carried out by senior scholars. I am a postdoctoral researcher on a fixed-term contract -- my career, and those of my peers, are undoubtedly going to be affected far more drastically by the adoption of LLMs in our community, than those of faculty members who have been in post for twenty years. This is even more true of current undergraduates and PhD students. The voices of early career researchers must therefore be heard in this conversation.

After eight months of near-daily use of agentic LLMs in my research, I remain sceptical. Use of these tools in a constrained environment, for `grunt work' such as debugging code does usually get such work done faster than I can do it myself. The insurmountable issue I have is the use of agentic LLMs to produce, in large part or entirely, a paper or journal article in which the results of a scientific analysis are reported and discussed. To do this, I believe, means the operator of the agentic system has a fundamental misunderstanding of what research \textit{is}.

According to the Frascati Manual developed by the OECD, research is ``creative and systematic work undertaken in order to increase the stock of knowledge'' \citep{Frascati}. For an activity to be classified as research, it must be ``novel, creative, uncertain, systematic, and transferable or reproducible''; without any one of these five elements, an activity is not research. Furthermore, the manual states that ``human input is inherent to creativity'' in research. I take this to be the definition of research, noting that human input is essential for an activity to qualify as research. In addition to creativity, a second, related aspect of research that demands human input is, in my opinion, interpretation; it is on this aspect of research, and the consequences of using agents to interpret data, that I focus on here.

The act of scientific interpretation is to take a result and communicate its meaning to the scientific community through a formal, peer-reviewed article. Data on their own are meaningless. For example, if an astronomer measured the value of the Hubble parameter, and posted their measurement on the arXiv preprint server as a single page PDF, containing their name, affiliation, and the number with its $1\sigma$ error bars, would we accept this as a reasonable contribution to the scientific literature?

The answer is no. This is because, even in the natural sciences, raw numbers are not sufficient to understand the world. It is the role of the researcher not only to create or collect data, but also to interpret it, and to communicate that interpretation to the community and to the general public. As scientists, we have a duty to impart meaning to the data we gather, to relate it to other measurements, to past knowledge, and to open ourselves to critique of those interpretations. In astronomy, at least, it is not the ends that matter, but rather the means; not the answer to a question, but the process by which we arrived at the answer. This viewpoint is rooted in postmodern hermeneutics \citep{Caputo2018}, but finds a familiar analogy in the aphorism attributed to George Box that all models are wrong, but some are useful \citep{Box1976}.

Agentic research misses this point entirely. Science without human interpretation does not exist. It may be objected that as long as humans remain `in the loop' with agentic systems, there is still some measure of interpretative ownership. I do not agree. Computation may be delegated to an agent because the correctness of its result is verifiable. Interpretation does not work in the same way, as there is no ground truth to compare against (all models are wrong), and to question the validity of an interpretation is to be conducting interpretation yourself. If the operator of an agentic system is capable of such interpretation, they should be the one doing it. On the other hand, if the operator of the agentic system is unable to interpret an output, they cannot honestly take ownership of or responsibility for either the output or its machine-interpretation, and the result is, by definition, something that is not research.

Furthermore, by allowing agentic LLMs to not only write code and perform calculations, but to also carry out their own hermeneutics on the result, we willingly give away what we as scientists once prided ourselves on -- our powers of deep, critical, and creative thought. Ingmar Bergman described the creative process thus: ``I throw a spear into the darkness. That is intuition. Then I must send an army into the darkness to find the spear. That is intellect.''~\citep{Bergman}. The act of using an agentic LLM for scientific interpretation is to throw a spear into the darkness with no hope of its retrieval, even with all the hordes of Babylon at one's disposal. 

What, then, will the end result of agentic `research' be? I think what we have to look forward to is a literature which is in some sense hollow -- produced without any human input, without human interpretation or understanding, at risk of collapsing under its own weight. The more these `hollow' papers are added to the literature, the less we can trust that literature. I call this phenomenon \textit{deliteration}; the process by which the astronomy literature will be gradually weakened by the proliferation of LLM-generated papers. Since we may consider that astronomy consists in its literature~\citep{DHogg2026}, agentic `research' thus signals the death knell for our collective understanding of the world. 

The endpoint of deliteration will be a climate of terrible distrust in astronomy, both of the literature itself and of our peers and colleagues. The only way to prevent the arrival of this gloomy future is for astronomy to undergo a serious restructuring, to alter the system that incentivises deliteration in the first place. Agentic AI is being sold to academics as a way to get ahead, to work faster, to be more productive than their colleague at the institution down the road who cannot afford to pay for the latest LLM. In a bizarre \textit{quid pro nilhilo}, we in turn are freely giving away our intellectual property to the tech companies as we continue to prompt their models.

Should we then give serious consideration to the ban-and-punish stance after all? Punish, certainly not. The interfaces with commercial LLMs are clearly designed to appeal to human psychology, and to encourage sustained use. Watching and waiting for the `thinking' icon in an agentic model's terminal interface to vanish and for new text to appear on the screen brings the same titillating thrill as the pull-to-refresh mechanism familiar to all social media users (itself reminiscent of casino slot machines). Besides, no scientist should be ashamed of being inquisitive about new technology. And the idea of attempting to implement a formal \textit{ban} on LLM contributions to the scientific literature engenders a feeling of futility only exceeded by the experience of watching the England men's team play World Cup football -- and would arguably encroach on academic freedom \citep{DHogg2026}.

Nevertheless, halting deliteration implies a future in which LLMs play no role in the interpretation of scientific results, and are not allowed to contribute to the literature, relegating their use to grunt work only. How can we achieve this? There are a number of immediately actionable steps: journals can require disclosure statements where an author must assert that all interpretative work was done by them rather than by an LLM; authors can make all data and results generated in the course of their investigation publicly available, so that others can understand how an interpretation was arrived at; and dedicated sessions can be held at workshops and conferences on the philosophy and ethics of AI use in astronomy. It is through precisely such interactions that my own views on this topic were informed.

In the longer term, we must tackle the `publish or perish' culture head-on. One practical avenue to change this culture would be to invest more time and money into the people who conduct research -- by and large, PhD students and postdocs. Ideally, if a temporary contract for them must be drawn up, it should be as lengthy as is possible -- four or five years at a minimum. Instead of being treated as disposable labour to be churned through as quickly as possible, students and postdocs should be given the space and time to learn the trade of astronomy and to make meaningful intellectual contributions at their own pace, with supportive supervision and mentoring. 

The notion of endless growth is a free market capitalist myth -- our community cannot sustain it. We must find a way to fix the system before deliteration seriously takes hold, or this may be one of the last articles you read that you can trust.

\section*{Acknowledgements}
The views expressed in this article were greatly enriched by discussions at the ``Agentic AI and research: opportunities and risks'' workshop held at SISSA in May 2026.

\section*{Competing interests}
The author declares no competing interests.

\bibliographystyle{unsrt}
\bibliography{agentic_research}

\end{document}